\documentclass[aps,onecolumn,a4paper,oneside,nobibnotes,nofootinbib,amsfonts,amssymb,amsmath,eqsecnum]{revtex4}
\usepackage{amssymb,amsmath} 
\usepackage{hyperref} 
\hypersetup{backref,colorlinks=true,linkcolor=blue,citecolor=red,linktocpage=true,breaklinks}
\textheight 9.5in \oddsidemargin -0.20in \evensidemargin 0.0in \newcommand{\sgn}{\operatorname{sgn}}
\DeclareMathOperator{\sech}{sech}
\begin{document}
\title{\bf Infinite families of (non)-Hermitian Hamiltonians
associated with exceptional $X_m$ Jacobi polynomials \vspace{.55 cm}} \author{Bikashkali Midya }
\email{bikash.midya@gmail.com} \author{Barnana Roy} \email{barnana@isical.ac.in} \affiliation{Physics \& Applied
Mathematics Unit, Indian Statistical
 Institute, Kolkata 700108, India.\vspace{1 cm}}

\begin{abstract} \noindent Using an appropriate change of variable, the Schr\"odinger equation
is transformed into a second-order differential equation satisfied by recently discovered Jacobi type $X_m$
exceptional orthogonal polynomials. This facilitates the derivation of infinite families of exactly solvable
Hermitian as well as non-Hermitian trigonometric Scarf potentials  and finite number of Hermitian and infinite number 
of non-Hermitian $\mathcal{PT}$-symmetric hyperbolic Scarf potentials. The bound state solutions of all these potentials
are associated with the aforesaid exceptional orthogonal polynomials. These infinite families of potentials are
shown to be extensions of the conventional trigonometric and hyperbolic Scarf potentials by the addition of some
rational terms characterized by the presence of classical Jacobi polynomials. All the members of a particular
family of these `rationally extended polynomial-dependent' potentials have the same energy spectrum and possess
translational shape invariant symmetry. The obtained non-Hermitian trigonometric Scarf potentials are shown
to be quasi-Hermitian in nature
ensuring the reality of the associated energy spectra. \end{abstract} 
\maketitle

\section{Introduction} \noindent In recent years there is a surge of interest in the study of exactly solvable quantum
systems associated with exceptional orthogonal polynomials introduced in the seminal paper by Gomez-Ullate et al
\cite{UKM10a,UKM10b}. The exceptional $X_m$ orthogonal polynomials (EOP) are the solutions of second-order
Sturm-Liouville eigenvalue problem with rational coefficients. A distinguishing property of these polynomials is that
the lowest eigen polynomial of the sequence need not be of degree zero, even though the full set of eigenfunctions
still forms a basis of the weighted $\mathcal{L}^2$ space.\\

Exceptional $X_1$ Laguerre or Jacobi type polynomials were shown to be eigenfunctions of rationally extended radial
oscillator or Scarf I potentials by the point canonical transformation method \cite{Qu08} and by supersymmetric
quantum mechanical method (SUSYQM) \cite{BQR09,Qu09}. The latter is based on a reparametrization of the starting
Hamiltonian and redefining the conventional superpotential in terms of modified couplings, while allowing for the
presence of extra rational contribution expressed in terms of some polynomial function. Depending on the degree of the
latter, the wave functions of the extended potentials are shown to be $X_1$ or $X_2$ Laguerre, Jacobi exceptional
orthogonal polynomials. Employing similar approach in the context of second order supersymmetry, rational extensions
of the  generalized Poschl-Teller, Scarf I potentials with the corresponding  bound state wavefunctions given in terms
of exceptional orthogonal polynomials have been constructed \cite{Qu09}. Construction of two distinct families of
Laguerre and Jacobi type $X_m$, $m=1,2,3...$, exceptional orthogonal polynomials as eigenfunctions of infinitely many
shape invariant potentials by deforming the radial oscillator and the Hyperbolic/Trigonometric Poschl Teller
potentials was made in refs.\cite{OS09,OS10a} and the properties of these $X_m$ exceptional orthogonal polynomials
have been thoroughly studied subsequently \cite{OS10c,HOS11,HS11,UKM12c,Qu11b}. The multi-indexed version
$X_{m_1,m_2,...,m_k}$ of exceptional $X_m$ orthogonal polynomials are constructed by using multi-step Darboux
transformation \cite{UKM12b}, Crum-Adler mechanism \cite{OS11b}, higher-order SUSYQM \cite{Qu11a} and multi-step
Darboux-Backlund transformation \cite{Gr12}.\\

$X_m$ exceptional orthogonal polynomials were shown to be obtainable through several other approaches such as
Darboux-Crum transformation \cite{STZ10,UKM10c,UKM12a},  Darboux-Backlund transformation \cite{Gr11a,Gr11c,Gr12} and
prepotential approach \cite{Ho11a,Ho11b}. In fact, $X_m$ Laguerre polynomials have been characterized in terms of an
isospectral Darboux transformation. The shape invariance of these polynomial families is shown to be a direct
consequence of the permutability property of the Darboux transformation \cite{UKM10c}. It is shown to be possible to
obtain rational extensions of every translationally shape invariant potentials via Darboux Backlund transformations
based on negative eigenfunctions built from excited states of the initial Hamiltonian. The potentials obtained by this method are either strictly isospectral or quasi isospectral extensions of
the initial ones and have enlarged shape invariance property. All the quantal systems related to the exceptional
Laguerre and Jacobi polynomials can be constructed by the prepotential approach \cite{Ho11a,Ho11b} which does not need
the concept of shape invariance and the Darboux-Crum transformation. The prepotential, the deforming function, the
potential, the eigenfunctions and eigenvalues can be derived within the same framework. The exceptional polynomials
obtained in this way are expressible as a bilinear combination of the deforming functions and its derivatives.\\

Recently, EOP have been studied in diverse scenarios. For instance, these new polynomials were shown to be associated
with the solutions of some conditionally exactly solvable potentials \cite{JR98,DR11}, solutions for position
dependent mass systems \cite{MR09}, main part of the eigenfunctions of the Dirac equations coupled minimally and
non-minimally with some external fields and the Fokker-Planck equations \cite{Ho11c}. Four types of infinitely many
exactly solvable Fokker-Planck equation which are the generalized, or deformed versions of the Rayleigh process and
the Jacobi process are also shown to related to EOP \cite{CH12}. The  Structure of the $X_m$ Laguerre polynomials was
considered within the quantum Hamilton-Jacobi formalism \cite{Ra+12}, within N-fold supersymmetry \cite{Ta10} and its
dynamical breaking in the context of position dependent mass scenario \cite{MRT12}. \\

On the other hand, relaxation of Hermiticity for the reality of the discrete spectrum of a quantum mechanical
Hamiltonian has given rise to some very interesting investigations in the last few years stepped up by a conjecture of
Bender and Boettcher \cite{BB98} that $\mathcal{PT}$-symmetric non-Hermitian Hamiltonians could possess real bound
state spectrum. The complex $\mathcal{PT}$-symmetric Hamiltonians possess a real discrete spectrum if the energy
eigenstates are also the eigenstates of $\mathcal{PT}$; otherwise, the $\mathcal{PT}$ symmetry is said to be
spontaneously broken and the energy eigenvalues occur in complex conjugate pairs \cite{Be05}. Here $\mathcal{P}$
stands for parity transformation and $\mathcal{T}$ stands for time reversal. Subsequently, Mostafazadeh \cite{Mo02}
showed that the reality of the spectrum of a $\mathcal{PT}$ symmetric Hamiltonian is ensured if the Hamiltonian
$\widetilde{H}$ is Hermitian with respect to a positive definite inner product $\langle.|\eta|.\rangle$ on the Hilbert
space $\mathcal{H}$ in which $\widetilde{H}$ is acting. This renders the Hamiltonian $\widetilde{H}$ to be
pseudo-Hermitian \cite{Mo10} $\widetilde{H}^{\dagger} = \eta \widetilde{H} \eta^{-1}$, where the Hermitian linear
automorphism $\eta : \mathcal{H} \rightarrow \mathcal{H}$ is bounded and positive definite.  A word of caution is due here if the 
operator $\eta$ is not a bounded operator. An unbounded metric operator $\eta$
 can not be used to define a consistent Hilbert space structure. In that case an alternative construction, which is not based on the 
 introduction of an $\eta$ operator, has been proposed in ref. \cite{KS04}. Another equivalent
condition for the reality of the energy spectrum of $\widetilde {H}$ is the quasi-Hermiticity \cite{KS04,SGH92,Mo04},
i.e. the existence of a invertible operator $\rho$ such that $H = \rho \widetilde{H} \rho^{-1}$ is Hermitian with
respect to usual inner product $\langle .,.\rangle$. Quasi-Hermitian Hamiltonian shares the same energy spectrum of
the equivalent Hermitian Hamiltonian $H$ and the wave functions are obtained by operating $\rho^{-1}$ on those of $H$.
Most of the analytically solvable non-Hermitian Hamiltonians are constructed by either making the coupling constant of
the potential present in the relevant Schr\"odinger Hamiltonian imaginary \cite{Ah01a,BR00} or by shifting the
coordinate with an imaginary constant \cite{LZ00,LZ01}. Several of these classes of Hamiltonians are argued to be
pseudo-Hermitian under $\eta = e^{-\alpha p}$ \cite{Ah01b,MRR10}.
 For a real $\alpha$ and $p=-i\frac{d}{dx}$, the operator $\eta$ shifts the coordinate $x$ to $x + i\alpha$.\\

In this paper we have found infinite families of exactly solvable Hermitian as well as non-Hermitian Hamiltonians
whose bound state wave functions are given in terms of exceptional $X_m$ Jacobi orthogonal polynomials. The non-Hermitian Hamiltonians, obtained here, are shown to be $\mathcal{PT}$-symmetric and quasi-Hermitian in nature rendering these Hamiltonians to
have real eigenvalues. Also, using SUSYQM, we have shown that all the infinite family of Hamiltonians have
translational shape invariant property. The motivation for doing this comes from the fact that compared to the exactly
solvable real potentials associated with exceptional orthogonal $X_m$ polynomials, a little attention
\cite{Mi12,BQR09} has been given to look for complex potentials having these polynomials as their bound state
eigenfunctions. In fact a few complex potentials have been found involving only $X_1$ type Jacobi, Laguerre
exceptional orthogonal polynomials. Considering the recent flurry of interest in non-Hermitian Hamiltonians with
$\mathcal{PT}$-symmetry and/or pseudo-Hermiticity the present study deserves to be worthwhile. The organization of the
paper is as follows: In section \ref{s2}, some properties of $X_m$ Jacobi EOPs, which are necessary to analyze the
results, are given. Construction of infinite families of Hermitian and non-Hermitian Hamiltonians by transforming the
relevant Schr\"odinger equation into the differential equation satisfied by the $X_m$ Jacobi polynomials are done in
section \ref{s3}. In section \ref{s4}, the shape invariance properties of the obtained potentials are discussed within
the framework of supersymmetric quantum mechanics. Finally, section \ref{s5} is devoted to the summary and outlook.

\section{Exceptional $X_m$ Jacobi Orthogonal Polynomials}\label{s2} \noindent We present here some properties of $X_m$
Jacobi polynomials which are studied in detail in ref. \cite{UKM12a}. For a fixed integer $m \ge 1$ and real $a, b >
-1$, the exceptional $X_m$ Jacobi orthogonal
 polynomials $\widehat{P}_n^{(a,b,m)}(x)$, $n=m, m+1, m+2...$, satisfies the differential equation
\begin{equation}
 (1-x^2) \frac{d^2y }{dx^2} + Q_1(x) \frac{dy}{dx} + R_1(x) y =0,\label{e22}
\end{equation} where $Q_1(x)$ and $R_1(x)$ can be written in terms of classical Jacobi orthogonal polynomial
$P_m^{(a,b)}(x)$, as \begin{equation}\begin{array}{ll} \displaystyle  Q_1(x) = (a-b-m+1) ~\frac{(1-x^2)
P_{m-1}^{(-a,b)}(x)}{P_m^{(-a-1,b-1)}(x)}-(a+1) (1+x) + (b+1)(1-x)\\ \\ \displaystyle  R_1(x) =
{b(a-b-m+1)}~\frac{(1-x) P_{m-1}^{(-a,b)}(x)}{P_m^{(-a-1,b-1)}(x)} + n^2 + n (a+ b- 2m+1) -2 b m.
\end{array}\label{e6} \end{equation}
 For $n,l \ge m$, this new family of polynomials are orthonormal with respect to the weight function
 $\widehat{W}_{a,b,m}(x) = \frac{(1-x)^a (1+x)^b}{P_m^{(-a-1,b-1)}(x)}$. Evidently, for $\mathcal{L}^2$-orthonormality
 to hold, the denominator of this weight function should be non-zero in $-1\le x \le 1$. According to the references
 \cite{Sz75,UKM12a} the polynomial $P_m^{(a+1,b-1)}(x)$ has no zeroes in $[-1,1]$ if and only if the following
 conditions hold simultaneously:
 \begin{equation}
 \begin{array}{ll}
(i)~ b \ne 0, ~~~ a, a-b-m+1 \notin \{0,1,...,m-1\},\\ (ii) ~a > m-2, ~~~ \sgn(a-m+1) = \sgn(b),
\end{array}\label{e17} \end{equation} where $\sgn(x)$ is the signum function. With these restrictions the
$\mathcal{L}^2$-norms of the exceptional $X_m$ Jacobi polynomials are given by \begin{equation}\begin{array}{ll}
\displaystyle \int^1_{-1} \frac{(1-x)^a(1+x)^b}{\left[P_m^{(-a-1,b-1)}(x)\right]^2} ~
\widehat{P}_n^{(a,b,m)}(x)~\widehat{P}_l^{(a,b,m)}(x) dx\\ \\ \displaystyle ~~~~~ = \frac{2^{a+b+1} (n+b)(n-2m +a+1)
\Gamma(n-m+a+2)\Gamma(n + b)}{(2n-2m+a+b+1)(n-m+a+1)^2 (n-m)!~
 \Gamma(n-m+a+b+1)} ~\delta_{nl},
 \end{array}\label{e21}
\end{equation} where $\delta_{nl}$ is the Kronecker delta function. Moreover, the $X_m$ Jacobi orthogonal polynomials
$\widehat{P}_n^{(a,b,m)}(x)$ are related to the classical Jacobi polynomials $P_n^{(a,b)}(x)$ by the following
relations \begin{subequations} \begin{equation} \widehat{P}_n^{(a,b,0)}(x) = P_n^{(a,b)}(x)\label{e16a} \end{equation}
\begin{equation} \begin{array}{ll}
 \displaystyle \widehat{P}_n^{(a,b,m)} = (-1)^m \left[\frac{a+b+j+1}{2(a+j+1)} (x-1)
 P_m^{(-a-1,b-1)}P_{j-1}^{(a+2,b)}\right.\\
 \displaystyle ~~~~~~~~~~~~~~~~~~~~~~~~~~~~~~~~~~~~~~~+
\left.\frac{a-m+1}{a+j+1} P_m^{(-a-2,b)}P_j^{(a+1,b-1)}\right],~~~~~ j=n-m \ge 0. \end{array}\label{e16b}
\end{equation} \end{subequations} It is worth mentioning here that following identities of classical Jacobi polynomial
are extensively used to simplify the results presented in the later part of the paper \begin{subequations}
\begin{alignat}{8} \! &(1-x^2) (a+b+m+2) P_{m-2}^{(a+2,b+2)} + 2 [b-a-(a+b+2) x] P_{m-1}^{(a+1,b+1)} + 4 m P_m^{(a,b)}
=0 \\ \! &  (x-1) (a + b + m + 1) P_{m-1}^{(a+1,b+1)} = 2 (a+m) P_m^{(a-1,b+1)} - 2 a P_m^{(a,b)}\\ \! &
P_m^{(a,b-1)} - P_m^{(a-1,b)} = P_{m-1}^{(a,b)}\\
 \! & \frac{d^r}{dx^r}P_m^{(a,b)} = \frac{\Gamma(a+b+m+r+l)}{2^r \Gamma(a+b+n+1)} P_{m-r}^{(a+r,b+r)}.
\end{alignat} \label{e12} \end{subequations} The identity (\ref{e12}a), which is absent in the handbook \cite{GR96},
directly follows from the differential equation of the classical Jacobi polynomials $P_m^{(a,b)}(x)$.

\section{Infinite family of Hermitian and non-Hermitian Hamiltonians associated with Jacobi type $X_m$
EOPs}\label{s3} \noindent Here we shall derive some exactly solvable Hermitian as well as non-Hermitian potentials
whose bound state solutions are connected to Jacobi type $X_m$ EOP. For this we have used an appropriate coordinate
transformation which transform the Schrodinger equation into a differential equation of $X_m$ Jacobi EOP. In the
following we first briefly describe this method.

\subsection{Construction method}\label{s3.1}
 \noindent  Following the discussion
of \cite{Le89,Mi12}, we consider transformation of the Schr\"{o}dinger equation (with $\hslash = 2 m = 1$)
\begin{equation}
 H \psi(x) = -\frac{d^2\psi(x)}{dx^2} + V(x) \psi(x) = E \psi(x).\label{e1}
\end{equation} into a second-order differential equation satisfied by a special function $F(g)$. For this the solution
$\psi(x)$ of the equation (\ref{e1}) can be assumed as \begin{equation}
 \psi(x) \sim f(x) ~ F(g(x)). \label{e8}
\end{equation} Substituting this form of $\psi(x)$ in (\ref{e1}), we obtain the following second-order ordinary
differential equation of the special function $F(g)$ \begin{equation}
 \frac{d^2 F}{dg^2} + Q(g) \frac{dF}{dg} + R(g) F(g) =0, \label{e2}
\end{equation} where $Q(g(x))$ and $R(g(x))$ are given by \begin{subequations}
  \begin{equation}
    Q(g) = \frac{g''}{g'^2} + \frac{2 f'}{f g'}, \label{e13}
  \end{equation}
  \begin{equation}
    R(g) = \frac{E-V(x)}{g'^2} + \frac{f''}{f g'^2},\label{e14}
  \end{equation}
\end{subequations} respectively. After doing some algebraic manipulations, the equations (\ref{e13}) and (\ref{e14})
reduce to \begin{subequations} \begin{equation}
 f(x) \approx g'(x)^{-1/2} ~ e^{^{\frac{1}{2} \int Q(g) dg}},\label{e7}
\end{equation} \begin{equation}
 E-V(x) = \frac{g'''}{2g'} - \frac{3}{4} \left(\frac{g''}{g'}\right)^2 + g'^2 \left(R-\frac{1}{2} \frac{dQ}{dg}
 -\frac{1}{4} Q^2\right).\label{e3}
\end{equation} \end{subequations} The above expression of $E-V(x)$ contains the unknown functions $Q(g)$, $R(g)$ and
the change of variable $g(x)$. A particular choice of the special function $F(g)$ provides the complete functional
forms of the first two unknowns, whereas a simple way of finding the unknown function $g(x)$ has been first proposed
by Bhattacharjee and Sudarshan \cite{BS62}. According to them, if there is a constant $(E)$ on the left-hand side of
(\ref{e3}), there must be one on the right-hand side too. Hence, a specific special function $F(g)$ and a reasonable
choice of $g(x)$ make the Schr\"odinger equation (\ref{e1}) an exactly solvable problem for the potential $V(x)$
(which is different for different choice of the special function). The solution of the corresponding Schr\"odinger
equation can be obtained using the equations (\ref{e8}) and (\ref{e7}) as \begin{equation} \psi(x) \sim g'(x)^{-1/2} ~
e^{^{\frac{1}{2} \int Q(g) dg}}~ F(g(x)).\label{e24} \end{equation} This procedure has been exploited in
ref.\cite{Le89} and a systematic list of shape invariant real potentials were obtained whose solutions are related to
the classical orthogonal polynomials. Analogous study, corresponding to $X_1$ EOPs, has been carried out in
\cite{Qu08} for Hermitian systems and in \cite{Mi12} for quasi-Hermitian systems. In the following we use the above
mentioned method to obtain infinitely many exactly solvable Hermitian as well as non-Hermitian Hamiltonians whose
bound state wave functions are connected with exceptional $X_m$ Jacobi orthogonal polynomials for arbitrary
$m=1,2,3...$.\\

Here, we choose the special function $F(g)$ to
be Jacobi type $X_m$ EOP e.g. $F(g) \sim \widehat{P}_n^{(a,b,m)}(g)$. In this case we have from equations (\ref{e22})
and (\ref{e2}) \begin{equation}
 Q(g) = \frac{Q_1(g)}{1-g^2}, ~~~~~~~ R(g) = \frac{R_1(g)}{1-g^2}, \label{e4}
\end{equation} where $Q_1(g)$ and $R_1(g)$ are given in equation (\ref{e6}). Using equations (\ref{e6}) and
(\ref{e4}), equation (\ref{e3}) reduces to
{\small
\begin{equation}\begin{array}{llll}
  \displaystyle E-V(x) = \frac{g'''}{2g'} - \frac{3}{4} \left(\frac{g''}{g'}\right)^2 +\frac{1-a^2}{4}
  \frac{g'^2}{(1-g)^2} +\frac{1-b^2}{4} \frac{g'^2}{(1+g)^2}\\\\
\displaystyle ~~~~~~~~~~~~~ + \frac{2 n^2+ 2 n (a + b - 2 m +1) + 2m(a-3b-m+1) + (a+1)(b+1)
}{2}\frac{g'^2}{1-g^2}\\\\
   \displaystyle ~~~~~~ + \frac{(a-b-m+1)[a+b+(a-b+1) g] g'^2}{1-g^2} ~\frac{ P_{m-1}^{(-a,b)}(g)}{
   P_{m}^{(-a-1,b-1)}(g)}- \frac{(a-b-m+1)^2 g'^2}{2}~\left[\frac{P_{m-1}^{(-a,b)}(g)}{P_m^{(-a-1,b-1)}(g)}\right]^2
.\label{e5} \end{array} \end{equation}}
At this point we choose $g'^2/(1-g^2) = \mbox{constant} = c\in
\mathbb{R}-\{0\}$, which is satisfied by
\begin{equation}
 g(x) = \sin (\sqrt{c}~ x + d).\label{e23}
\end{equation} Without loss of generality we will first choose $d=0$. Nonzero real values of $d$ do not make any
significant difference in the potential and its solutions. The nonzero complex values of $d$ give rise to complex
potentials with real spectra as will be shown later. The choice (\ref{e23}) of $g(x)$ helps to generate a constant
term in the right hand side corresponding to the energy in the left hand side of equation (\ref{e5}). Here two cases
may arises,
 depending on the sign of the chosen constant $c$.
 In the following, we consider two cases $c >$ and $<0$ separately. $c>0$ gives the infinite numbers of
rationally extended Hermitian as well as quasi-Hermitian trigonometric Scarf potentials while $c<0$ corresponds 
to the finite number of Hermitian and infinite number of non-Hermitian $\mathcal{PT}$-symmetric hyperbolic Scarf potential family.

\subsection{Infinite family of Hermitian Hamiltonians}\label{s3.2}
\noindent {\bf Case I: \underline{$c > 0, d=0$;  extended real trigonometric Scarf potential family}}

Let $c=k^2$,
$k\ne0$. In this case, we have $g(x) = \sin k x$. Substituting $g(x) = \sin kx$ in equation (\ref{e5}) and separating
out the potential $V(x)$ and energy $E$, we have \begin{subequations} \begin{equation} \begin{array}{llll}
 V^{(m)}(x) = \frac{k^2 (2a^2+ 2b^2 - 1)}{4} \sec^2 kx -\frac{k^2 (b^2-a^2)}{2} \sec kx \tan kx - 2 k^2 m (a-b-m+1)\\
\displaystyle ~~~~~~~~~~~~ - k^2(a-b-m+1)[a+b+(a-b+1) \sin kx] ~\frac{ P_{m-1}^{(-a,b)}(\sin kx)}{
P_{m}^{(-a-1,b-1)}(\sin kx)}\\\\ \displaystyle ~~~~~~~~~ + \frac{k^2 (a-b-m+1)^2 \cos^2 kx}{2} \left[\frac{
P_{m-1}^{(-a,b)}(\sin kx)}{P_{m}^{(-a-1,b-1)}(\sin kx)}\right]^2 ~, ~~~~ -\frac{\pi}{2 k} < x < \frac{\pi}{2k}
\end{array}\label{e10a} \end{equation} \begin{equation} \hspace{-4.5cm} E_n^{(m)} = \frac{k^2}{4} (2 n -2 m + a +
b+1)^2, ~~~~ n = m,m+1, m+2,...\label{e10b} \end{equation} \end{subequations} Also, the expression of the wave
function $\psi(x)$ given in equation (\ref{e24}) reduces to \begin{equation}
 \psi_n^{(m)}(x) = {\cal{N}}^{(m)}_n \frac{(1-\sin kx)^{\frac{a}{2}+\frac{1}{4}} (1+\sin
 kx)^{\frac{b}{2}+\frac{1}{4}}}
{P_m^{(-a-1,b-1)}(\sin kx)} ~ \widehat{P}_n^{(a,b,m)}(\sin kx), ~~~~n\ge m \label{e15} \end{equation} The
normalization constant $\mathcal{N}_n^{(m)}$ can be obtained using equation (\ref{e21}), in the following form
\begin{equation} {\cal{N}}_n^{(m)} =  \left[\frac{k~(2n-2m+a+b+1)(n-m+a+1)^2 ~(n-m)!~ \Gamma(n-m+a+b+1)}{2^{a+b+1}
(n+b)(n-2m +a+1) \Gamma(n-m+a+2)\Gamma(n +b)}\right]^{\frac{1}{2}} \end{equation} At this point few remarks on the
obtained potentials $V^{(m)}(x)$ are worth to be mentioned.
 \begin{itemize}
 \item[$\bullet$]{The new potentials $V^{(m)}(x)$, $m=0,1,2...$ become singular at the zeroes of the Jacobi
     polynomial $P_m^{(-a-1,b-1)}(x)$ in the interval $(-\frac{\pi}{2k},\frac{\pi}{2k})$. These singularities can
     be avoided by restricting the potential parameters $a,b$ according to the conditions mentioned in equation
     (\ref{e17}).}
     \item[$\bullet$]{The potentials $V^{(m)}(x)$ given in (\ref{e10a}) are infinite in number because each
         integer values of $m \ge 0$ gives rise to a new exactly solvable potential.}
     \item[$\bullet$]{For $m=0$ the last three terms in the expression of $V^{(m)}(x)$ vanish and one is left with
         the well known trigonometric Scarf potential $V^{(0)}(x) = \frac{k^2 (2a^2+ 2b^2 - 1)}{4} \sec^2 kx
         -\frac{k^2 (b^2-a^2)}{2} \sec kx \tan kx$. It's solutions are given by classical Jacobi polynomials
         $P_m^{(a,b)}(x)$ as can be easily verified using equations (\ref{e15}) and (\ref{e16a}). All other
         choices of $m$ make the potentials $V^{(m)}(x)$ as the extensions of $V^{(0)}(x)$ by addition of some
         rational terms which are explicitly depends on the classical Jacobi polynomials $P_m^{(a,b)}(x)$.
         Therefore we can interpret $V^{(m)}(x)$ as the rationally extended polynomial-depended trigonometric
         Scarf potential family.}
         \item[$\bullet$]{All the members $V^{(1)}(x), V^{(2)}(x), V^{(3)}(x)...$ of the polynomial-depended
             trigonometric Scarf potential family $V^{(m)}(x)$ are isospectral i.e. they share the same energy
             spectrum $E_{n'}^{(m)} =  \frac{k^2}{4} (2 n' + a + b+1)^2$, $n'=0,1,2...$, where we have reset the
             quantum number $n=n'+m$.}
 \end{itemize}

\noindent In the following we are giving some simple cases of the results obtained in equations (\ref{e10a}),
(\ref{e10b}) and (\ref{e15}).\\

\noindent { (a) \underline{$m=0$}:} For $m=0$, we have from equations (\ref{e10a}) and (\ref{e10b})
\begin{equation}\begin{array}{ll} \displaystyle V^{(0)}(x) = \frac{k^2(2 a^2 + 2 b^2-1)}{4} \sec^2 kx - \frac{k^2(b^2
- a^2)}{2} \sec kx \tan kx\\ \\ \displaystyle E= \frac{k^2}{4}(2n +a+b+1)^2
 \end{array}\label{e11}
\end{equation} By redefining the parameters $a = \alpha - \beta - 1/2, ~~ b = \alpha + \beta - 1/2$, the above
equation (\ref{e11}) reduces to the well known trigonometric Scarf potential \cite{Le89,CKS00} and its bound state
energy spectrum given by \begin{equation}
 \begin{array}{ll}
   \displaystyle V^{(0)}(x) = k^2\left[\alpha(\alpha-1)+\beta^2\right] \sec^2 kx - k^2 \beta(2\alpha-1) \sec kx \tan
   kx\\ \\
\displaystyle E_n^{(0)}= k^2(n + \alpha)^2, ~~~~~~ n=0,1, 2...
 \end{array}
\end{equation} respectively.  Corresponding wave functions can be written in terms of classical Jacobi polynomials
\begin{equation}
 \psi^{(0)}_n(x) = \mathcal{N}_n^{(0)} (1-\sin kx)^{(\frac{\alpha-\beta}{2})}(1+\sin kx)^{(\frac{\alpha+\beta}{2})}
 ~P_n^{(\alpha-\beta-
\frac{1}{2},\alpha+\beta-\frac{1}{2})} (\sin kx) \end{equation}

\noindent {(b) \underline{$m = 1$}}: For $m=1$ and the same choices of parameters mentioned in the case of $m=0$,
equation (\ref{e5}) reduces to \begin{equation}
 \begin{array}{ll}
  V^{(1)}(x)=k^2\left[\alpha(\alpha-1)+\beta^2\right] \sec^2 kx - k^2 \beta(2\alpha-1) \sec kx \tan kx + \frac{2 k^2 (
  2 \alpha-1)}{2 \alpha - 1 - 2 \beta \sin k x} - \frac{2 k^2 \left[(2 \alpha-1)^2 - 4 \beta^2\right]}{(2 \alpha - 1 -
  2 \beta \sin k x)^2 }\\ \\
E_n^{(1)} = k^2 (n+\alpha-1)^2, ~~~~~~~~ n=1, 2, 3...
 \end{array}
\end{equation} The potential $V^{(1)}(x)$ is the rationally extended trigonometric Scarf potential, studied earlier in
ref. \cite{Qu08}. The bound state solutions can be written in terms of exceptional $X_1$ Jacobi polynomials
\cite{UKM10a,UKM10b} \begin{equation}
 \psi^{(1)}_n(x) = \mathcal{N}_n^{(1)} \frac{(1-\sin kx)^{(\frac{\alpha-\beta}{2})}(1+\sin
 kx)^{(\frac{\alpha+\beta}{2})}}{2 \alpha-1-2\beta \sin kx} ~
\widehat{P}_{n}^{(\alpha-\beta-\frac{1}{2},\alpha+\beta-\frac{1}{2},1)} (\sin kx), ~~~ n=1,2,... \end{equation}

\noindent {(c) \underline{$m=2$}}: In this case we have the following potential, energies and bound state wave
functions involving $X_2$ exceptional Jacobi orthogonal polynomials {\begin{equation}
 \begin{array}{llll}
   V^{(2)}(x)= k^2\left[\alpha(\alpha-1)+\beta^2\right] \sec^2 kx - k^2 \beta(2\alpha-1) \sec kx \tan kx\\ \\
 \displaystyle ~+\frac{4 k^2[ 3 ( 2 \alpha-1 ) ( 2 \beta +1) \sin k x - 2\beta (2\beta+1)-8\alpha(\alpha-1)]}
{2(\beta+1)(2\beta+1)\sin^2 k x +2(2\beta+1)(2 \alpha-1) \sin kx  + 4\alpha (\alpha-1)- 2 \beta -1  }\\ \\
\displaystyle  + \frac{8 (2 \beta+1)^2 k^2 \cos^2 k x[2 (1 + \beta) \sin k x - 2 \alpha + 1]^2}
{[2(\beta+1)(2\beta+1)\sin^2 k x +2(2\beta+1)(2 \alpha-1) \sin kx  + 4\alpha (\alpha-1)- 2 \beta -1]^2} - 8 k^2\\ \\

E_n^{(2)} = \frac{k^2}{4} (n+\alpha -2)^2, ~~~~~~~~ n = 2, 3, 4... \end{array} \end{equation} and \begin{equation}
\begin{array}{ll}
 \displaystyle \psi_n^{(2)} (x) = \mathcal{N}_n^{(2)} \frac{(1-\sin kx)^{(\frac{\alpha-\beta}{2})}(1+\sin
 kx)^{(\frac{\alpha+\beta}{2})}}{2(\beta+1)(2\beta+1)\sin^2 k x +2(2\beta+1)(2 \alpha-1) \sin kx  + 4\alpha
 (\alpha-1)- 2 \beta -1}\\\\
 \displaystyle ~~~~~~~~~~~~~~~~~~~~~~\times \widehat{P}_{n}^{(\alpha-\beta-\frac{1}{2},\alpha+\beta-\frac{1}{2},2)}
 (\sin kx), ~~~ n=2,3,...
\end{array} \end{equation} respectively.

\subsection{Infinite family of non-Hermitian Hamiltonians}\label{s3.3}
\noindent {\bf Case II: \underline{$c>0, d=i\epsilon$; extended quasi-Hermitian trigonometric Scarf potential family}}

In the previous section \ref{s3.2}, we have
considered the constant of integration $d$ to be zero. The nonzero value of $d$ corresponds to the shift of the
coordinate $x$. In the Hermitian case this co-ordinate shift is not so important as they do not influence the energy
eigenfunctions and eigenvalues. However, in $\mathcal{PT}$-symmetric quantum mechanics imaginary coordinate shift
plays a significant role. This gives rise to complex potentials with entirely real spectrum.
 Here we shall examine how the potentials $V^{(m)}(x)$ and associated energy spectra
behave if we allow purely imaginary value of $d$. For this, we set $d=i \epsilon$ and $c = k^2$, in which case the
equations (\ref{e23}) and (\ref{e10a}) reduce to \begin{equation}
  g(x) = \sin (kx + i \epsilon)
\end{equation} and \begin{equation}
 \begin{array}{llll}
 \widetilde{V}^{(m)}(x) = \frac{k^2 (2a^2+ 2b^{2} - 1)}{4} \sec^2 (kx + i \epsilon) -\frac{k^2 (b^{2}-a^{2})}{2} \sec
 (kx + i \epsilon) \tan (kx + i \epsilon) - 2 k^2 m (a-b-m+1)\\\\
\displaystyle ~~~~~~~~ - k^2(a-b-m+1)[a+b+(a-b+1) \sin (kx + i \epsilon)] ~\frac{ P_{m-1}^{(-a,b)}(\sin (kx + i
\epsilon))}{ P_{m}^{(-a-1,b-1)}(\sin (kx + i \epsilon)}\\\\ \displaystyle ~~~~~~~~~~~+ \frac{k^2 (a-b-m+1)^2 \cos^2
(kx + i \epsilon)}{2} \left[\frac{ P_{m-1}^{(-a,b)}(\sin (kx + i \epsilon))}{P_{m}^{(-a-1,b-1)}(\sin (kx + i
\epsilon))}\right]^2 ~, ~~~ -\infty <x < \infty\\ \\
 \widetilde{E}_n^{(m)} = \frac{k^2}{4} (2 n -2 m + a + b +1)^2, ~~~~ n = m,m+1, m+2,...
\end{array} \label{e18} \end{equation} Like the Hermitian case, the potentials $\widetilde{V}^{(m)}(x)$, $m=0,1,2,..$
are infinite in number and all the members are isospectral. Here, $a,b$ are chosen to be real otherwise the solution
of this potential will contain Jacobi polynomials with complex indices and complex arguments. Such complex polynomials
are not suitable for physical applications because of their non-trivial orthogonality properties which depend on the
interplay between integration contour and parameter values. It is to be noted here that the non-Hermitian Hamiltonians
$\widetilde{H}^{(m)}$ with the potentials $\widetilde{V}^{(m)}(x)$ have entirely real bound state energies
$\widetilde{E}_n^{(m)}$. This can be proved in the following way. Let us define a Hermitian, positive definite
operator $\rho$ as \begin{equation} \rho = e^{\frac{\epsilon}{k} p}, ~~~p= -i \frac{d}{dx} \label{e60} \end{equation}
which has the following properties \cite{Ah01b} \begin{equation} \rho x \rho^{-1} = x-\frac{i \epsilon}{k},~~~~ \rho p
\rho^{-1} = p, ~~~~\rho f(x) \rho^{-1} = f\left(x-\frac{i \epsilon}{k}\right). \end{equation} With the help of the
operator $\rho$, we have the following similarity transformation \begin{equation} \rho \widetilde{V}^{(m)}(x)
\rho^{-1} = V^{(m)}(x) \end{equation} This shows that the non-Hermitian Hamiltonians $\widetilde{H}^{(m)}$ are
quasi-Hermitian with respect to the Hermitian, positive definite operator $\rho$. The equivalent Hermitian
counterparts are the Hamiltonians  with the potentials $V^{(m)}(x)$ (\ref{e10a}) corresponding to $d=\epsilon=0$. The
energy eigenfunctions of the potentials $\widetilde{V}^{(m)}(x)$ can be obtained using $\widetilde{\psi}^{(m)}_n =
\rho^{-1} \psi^{(m)}_n(x) = \psi^{(m)}_n (x+ i \epsilon)$, where $\psi^{(m)}_n(x)$ are given by equation (\ref{e15}).
It will not be difficult to show that the Hamiltonian $\widetilde{H}^{(m)}$ is also pseudo-Hermitian i.e.
\begin{equation} \eta \widetilde{H}^{(m)} \eta^{-1} = \widetilde{H}^{(m)\dag} \end{equation} with respect to the
positive definite operator \begin{equation} \eta = \rho^2 = e^{\frac{2 \epsilon}{k} p}.\label{e61} \end{equation}
$\mathcal{PT}$-symmetry of the potential $\widetilde{V}^{(m)}(x)$ cannot be achieved, in general, if $d$ has a
non-zero real component, because the finite shift along the coordinate $x$ makes the potential different from its
$\mathcal{PT}$ counterpart. Now, we shall try to determine under what condition the above potential remains
$\mathcal{PT}$ invariant. A necessary requirement for a Hamiltonian to be $\mathcal{PT}$-symmetric is that the
potential should satisfy following condition \begin{equation}
 \widetilde{V}^{(m)*}(-x) = \widetilde{V}^{(m)}(x).
\end{equation}
 This implies that for $m=0,1$ the potentials $\widetilde{V}^{(m)}(x)$ are $\mathcal{PT}$-symmetric 
 iff $a = \pm b$. For $m\ge 2$, they are $\mathcal{PT}$-symmetric if $a=-b$. Hence, we have derived infinitely 
 many non-Hermitian rationally extended trigonometric Scraf potentials with entirely
real spectra and whose solutions are associated with exceptional $X_m$ Jacobi EOP.\\

\noindent {\bf Case III: \underline{ $c < 0, d=0$; non-Hermitian $\mathcal{PT}$-symmetric hyperbolic Scarf potential family}}

Let $c=-k^2$, $k\ne0$ and $d=0$. In this case we have from (\ref{e23}) $g(x) = {i} \sinh k x$. The corresponding potential in equation (\ref{e5}) (we rename them as
$U^{(m)}(x)$) reduces to
\begin{equation} \begin{array}{lll}
 U^{(m)}(x) = -\frac{k^2 (2a^2+ 2b^2 - 1)}{4} \sech^2 kx + i \frac{k^2 (b^2-a^2)}{2} \sech kx \tanh kx + 2 k^2 m
 (a-b-m+1)\\\\
\displaystyle ~~~~~~~~ + k^2(a-b-m+1)[a+b + (a-b+1) i \sinh kx] ~\frac{ P_{m-1}^{(-a,b)}(i \sinh kx)}{
P_{m}^{(-a-1,b-1)}(i \sinh kx)}\\\\ \displaystyle ~~~~~~~~~ - \frac{k^2 (a-b-m+1)^2 \cosh^2 kx}{2}
\left[\frac{P_{m-1}^{(-a,b)}(i \sinh kx)}{P_{m}^{(-a-1,b-1)}(i \sinh kx)}\right]^2 ~, ~~~~~~ -\infty < x < \infty
\end{array}\label{e25} \end{equation}
The bound state wave functions and energy spectrum of the Schr\"odinger equation
for this potential $U^{(m)}(x)$ are given by equations (\ref{e5}) and (\ref{e24}), as \begin{subequations} \begin{equation}
  \psi_n^{(m)}(x) = {\cal{N}}^{(m)}_n \frac{(1-i \sinh kx)^{\frac{a}{2}+\frac{1}{4}} (1+ i \sinh
  kx)^{\frac{b}{2}+\frac{1}{4}}}{P_m^{(-a-1,b-1)}(i \sinh kx)} ~ \widehat{P}_n^{(a,b,m)}(i \sinh kx),
\end{equation} \begin{equation}
 E_n^{(m)} = -\frac{k^2}{4} (2 n -2 m + a + b+1)^2, ~~~ n = m,m+1,m+2,...< \left(m-\frac{a+b+1}{2}\right) \\
\end{equation} \end{subequations} respectively.  Few comments on the potential family $U^{(m)}(x)$ are as follows:
\begin{itemize}
\item{Since the potential
is free from singularity in $-\infty < x < \infty$ so the wave function should vanish asymptotically at $\pm \infty$. This imposes an upper bound on the quantum number $n < \left( m -\frac{a+b+1}{2}\right).$}

\item{The potentials $U^{(m)}(x)$, $m=0,1,2...$ are in general non-Hermitian with entirely real energy spectrum. Moreover, 
they are $\mathcal{PT}$-symmetric for all real values of $a,b$. The potentials $U^{(m)}(x)$ are also rationally extended version of the conventional non-Hermitian hyperbolic Scarf potentials \cite{Le02} by addition of some polynomial dependent terms.}

\item{ For $m=0,1$, the corresponding potentials $U^{(m)}(x)$ are real provided $a=b$. For $m \ge 2$, they become real if $a=-b$. But in the latter case the corresponding energy spectrum becomes empty. So the real potential corresponding to $m\ge2$ are not physically interesting.}

\item{ All the rationally
extended non-Hermitian hyperbolic Scarf potentials are isospectral with energy spectrum $E_{n'}^{(m)} = -\frac{k^2}{4} (2 n' + a +
b+1)^2$, $n'=0,1,2...$}
\end{itemize}
Hence, the potentials $U^{(m)}(x), m=0,1,2...$ can be interpreted as infinite family of non-Hermitian 
$\mathcal{PT}$-symmetric extended hyperbolic Scarf potentials with entirely real energy spectrum. In particular case $a=b$ the 
potentials $U^{(0)}$ and $U^{(1)}(x)$ become Hermitian.

\section{Supersymmetric Shape Invariance Approach}\label{s4} \noindent Here, we combine the results, obtained in
section \ref{s3}, with supersymmetric quantum mechanics. Supersymmetry makes use of two linear first-order
differential operators $A^{(m)\pm} = \mp \frac{d}{dx} + W^{(m)}(x) $, where $W^{(m)}(x)$ is the superpotential and
generally defined \cite{CKS00} in terms of the ground state wave function $\psi^{(m)}_m(x)$ as \begin{equation}
 \psi^{(m)}_m(x) \sim \exp \left(-\int^x W^{(m)}(r) dr \right) \label{e62}
\end{equation} The two operators $A^{(m)\pm}$ give rise to two partner Hamiltonians $H^{(m)\pm}$ \begin{equation}
 H^{(m)\mp} = A^{(m)\pm} A^{(m)\mp} -\frac{d^2}{dx^2} + V^{(m)\mp}(x) - \varepsilon,
\end{equation} where $\varepsilon$ is the factorization energy and $V^{(m)\pm}$ are the two partner potentials given
in terms of superpotential \begin{equation}
 V^{(m)\pm} (x) = W^{(m)}(x)^2 \pm W^{(m)'}(x).\label{e51}
\end{equation} The partner Hamiltonians have the same spectrum except the zero energy state i.e.
 $E_{n+1}^{(m)-} = E_{n}^{(m)+}$ and $E_0^{(m)-} = 0$.
If the zero energy state of one of the Hamiltonians $H^{(m)\pm}$ is known, the eigenfunctions of the other Hamiltonian
can be determined using $\psi_n^{(m)+} \sim A^- \psi^{(m)-}_{n+1}$ and $\psi_n^{(m)-} \sim A^+ \psi^{(m)+}_{n-1}$. The
two partner potentials $V^{(m)\pm}(x)$ are said to be shape invariant if they satisfy \cite{Ge83} \begin{equation}
V^{(m)+} (x,a_0) = V^{(m)-} (x,a_1) + R(a_0) \end{equation} where $a_1 = f(a_0)$ and $R(a_0)$ is independent of $x$.
The beauty of shape invariance property is that whenever two supersymmetric partner potentials are related by shape
invariance condition, the energy eigenvalues and the eigenvectors can be determined algebraically \cite{CKS00}.

For convenience we identify the potential $V^{(m)}(x)$, given in equation (\ref{e10a}), with $V^{(m)-}(x)$. The ground
state solution [follow the equation (\ref{e15})] of this potential can be written as \begin{equation} \psi_m^{(m)-}
\sim (1-\sin kx)^{\frac{a}{2}+\frac{1}{4}} (1+\sin kx)^{\frac{b}{2}+\frac{1}{4}} \left[ 1-
\frac{P_{m-1}^{(-a-1,b)}(\sin k x)}{P_m^{(-a-1,b-1)}(\sin k x)}\right]. \label{e63} \end{equation} Using eqns.
(\ref{e62}), (\ref{e63}) and after some algebraic manipulations with the help of the recurrence relations (\ref{e12})
we obtain the superpotential as \begin{equation}\begin{array}{lll}
 \displaystyle W^{(m)}(x) = \frac{k(a-b)}{2} \sec k x + \frac{k(a+b+1)}{2} \tan k x \\
\displaystyle ~~~~~ ~~~~~~~~~~~~~~- \frac{k(a-b-m+1)\cos k x}{2}\left[\frac{P_{m-1}^{(-a,b)}(\sin
kx)}{P_m^{(-a-1,b-1)}(\sin kx)} - \frac{P_{m-1}^{(-a-1,b+1)}(\sin kx)}{P_m^{(-a-2,b)}(\sin kx)}\right] \end{array}
\label{e64} \end{equation} From equations (\ref{e51}) and (\ref{e64}) we get the simplified expressions of the partner
potentials $V^{(m)\mp}$ as \begin{equation}
 \begin{array}{lll}
V^{(m)-}(a,b,x) = \frac{k^2 (2a^2+ 2b^2 - 1)}{4} \sec^2 kx -\frac{k^2 (b^2-a^2)}{2} \sec kx \tan kx - 2 k^2 m
(a-b-m+1)\\ \displaystyle ~~~~~~~~~ - k^2(a-b-m+1)[a+b+(a-b+1) \sin kx] ~\frac{ P_{m-1}^{(-a,b)}(\sin kx)}{
P_{m}^{(-a-1,b-1)}(\sin kx)}\\ \displaystyle ~~~~~~~~~ + \frac{k^2 (a-b-m+1)^2 \cos^2 kx}{2} \left(\frac{
P_{m-1}^{(-a,b)}(\sin kx)}{P_{m}^{(-a-1,b-1)}(\sin kx)}\right)^2 - \frac{k^2(a+b+1)^2}{4}
   \end{array}\label{e19}
\end{equation} and \begin{equation}
 \begin{array}{lll}
 \displaystyle  V^{(m)+}(a,b,x) = \frac{k^2 \left[2(a+1)^2+ 2(b+1)^2 - 1\right]}{4} \sec^2 kx -\frac{k^2
 \left[(b+1)^2-(a+1)^2\right]}{2} \sec kx \tan kx\\
\displaystyle ~~~~~~ ~- k^2(a-b-m+1)[a+b+2+(a-b+1) \sin kx] ~\frac{ P_{m-1}^{(-a-1,b+1)}(\sin kx)}{
P_{m}^{(-a-2,b)}(\sin kx)}\\ \displaystyle ~~~~~~~+ \frac{k^2 (a-b-m+1)^2 \cos^2 kx}{2} \left(\frac{
P_{m-1}^{(-a-1,b+1)}(\sin kx)}{P_{m}^{(-a-2,b)}(\sin kx)}\right)^2 - 2 k^2 m (a-b-m+1)-  \frac{k^2(a+b+1)^2}{4}
 \end{array}\label{e20}
\end{equation} The potential $V^{(m)-}$ matches, apart from an additive factorization energy $\varepsilon=
\frac{k^2(a+b+1)^2}{4}$, with the potential (\ref{e10a}) obtained in section \ref{s3}. The potential $V^{(m)}(x)$ is
another infinite set of exactly solvable potential. It is not very difficult to show that the bound state solutions of
this potentials are also associated with $X_m$ Jacobi Polynomials. Also, it is very easy to check that the two partner
potentials $V^{(m)\mp}$ given in equations (\ref{e19}) and (\ref{e20}) are connected to each other by
\begin{equation}
 V^{(m)+}(a,b,x) = V^{(m)-}(a+1,b+1,x) + k^2 (a+b+2)
\end{equation} i.e. they have the translational shape invariance symmetry. Analogously one can study other generalized
potential families, obtained in section \ref{s3}, in the framework of supersymmetric quantum mechanics to show that
they have shape invariance symmetry.

\section{Summary}\label{s5} \noindent  In summary, we have obtained infinitely many exactly solvable Hermitian as well
as non-Hermitian trigonometric Scarf potentials $V^{(m)}(x)$ and $\tilde{V}^{(m)}(x), m=0,1,2...$ respectively. We have also obtained finite number of Hermitian and infinite number of non-Hermitian $\mathcal{PT}$-symmetric hyperbolic Scarf potentials $U^{(m)}(x)$ with entirely real energy spectra. The bound state wave functions of all these potentials are associated with the exceptional $X_m$ Jacobi
polynomials. All the potentials belonging to a particular family of potentials $V^{(m)}(x)$, or $U^{(m)}(x)$, or $\widetilde{V}^{(m)}(x)$ are isospectral to each other. The supersymmetric partners of the potentials, obtained here, possess shape invariant symmetry. The non-Hermitian Hamiltonians involving the complex trigonometric Scarf potentials $\tilde{V}^{(m)}(x)$ are shown to be quasi-Hermitian with respect to a invertible
operator $\rho = e^{\frac{\epsilon}{k}p}$. This implies that corresponding energy spectra are all real. It has also been shown that for $a=-b$ the potentials $\widetilde{V}^{(m)}(x)$ are $\mathcal{PT}$-symmetric whereas the potentials
${U}^{(m)}(x)$ are $\mathcal{PT}$-symmetric for all $a, b$.

\section*{Acknowledgment} \noindent One of the authors (B.M.) thanks Robert Milson for communication regarding
exceptional $X_m$ Jacobi Polynomials.

 \end{document}